\documentclass[aps,showpacs,showkeys,preprintnumbers]{revtex4}

\usepackage{graphicx}
\usepackage{latexsym}
\usepackage{amssymb}

\def\beq{\begin{equation}}
\def\eeq{\end{equation}}
\def\rmd{{\rm d}} 

\begin{document}

\title{Scattering of particles by radiation fields: a comparative analysis}

\author{D. Bini$^{a,b}$, A. Geralico$^{c,b}$, M. Haney$^{c,b}$ and R. T. Jantzen$^{d,b}$}
\affiliation{
${}^a$Istituto per le Applicazioni del Calcolo ``M. Picone,'' CNR, I-00185 Rome, Italy\\
${}^b$ICRA, University of Rome ``La Sapienza,'' I-00185 Rome, Italy\\
${}^c$Physics Department, University of Rome ``La Sapienza,'' I-00185 Rome, Italy\\
${}^d$Department of Mathematical Sciences, Villanova University, Villanova, PA 19085, USA
}

\begin{abstract}
The features of the scattering of massive 
neutral
particles propagating in the field of a gravitational plane wave are compared with those characterizing their interaction with an electromagnetic radiation field. 
The motion is geodesic in the former case, whereas in the case of an electromagnetic pulse it is accelerated by the radiation field filling the associated spacetime region. 
The interaction with the radiation field is modeled by a force term entering the equations of motion proportional to the 4-momentum density of radiation observed in the particle's rest frame.
The corresponding classical scattering cross sections are evaluated too. 
\end{abstract}



\maketitle

\section{Introduction}

The production of gravitational waves as well as of electromagnetic pulses is expected to occur in many violent astrophysical processes, like the merging of compact binaries and high energy phenomena involving strong magnetic fields and accelerating sources of the electromagnetic field.
Gravitational and electromagnetic waves are also believed to interact in a variety of ways. 
There are many exact solutions of Einstein's field equations that describe colliding plane gravitational and electromagnetic waves on a flat Minkowski background \cite{grif3}.
Furthermore, several studies in the literature have shown how gravitational radiation affects the propagation of electromagnetic signals by modifying their direction, amplitude, wavelength and polarization either in vacuum or in the presence of conductive plasmas, leading also to the possibility of resonances between gravitational and electromagnetic sources which could be used either as more efficient gravity-wave detection methods or as a general relativistic
mechanism of amplifying large-scale magnetic fields (see, e.g., Ref.~\cite{tsagas} and references therein).

The scattering of massive and massless neutral scalar particles by plane gravitational waves has been investigated both in the classical and quantum regime by Garriga and Verdaguer \cite{garriga}.  
They also defined the classical cross section for scattering of geodesic particles in the case in which the wave region is sandwiched between two flat spacetime regions. 
The propagation of a test electromagnetic field on the background of an exact gravitational plane wave with single polarization has been recently investigated in Ref.~\cite{bfho}. 
It has been shown there that the physical effects due to the exact gravitational wave on the electromagnetic field, i.e., phase shift, change of the polarization vector, angular deflection and delay of photon beams in a Michelson interferometer, could be measured by various detection methods.

An electromagnetic wave propagating over a spacetime region makes it not empty and not flat. 
Therefore, the spacetime curvature associated with an electromagnetic pulse, namely the associated gravitational field, induces observable effects on test particle motion.
Unlike the case of a plane gravitational wave the resulting motion will no longer be geodesic, but massive particles will be accelerated by the radiation field filling the associated spacetime region. 
The features of test particle motion in the gravitational field associated with an electromagnetic plane wave has been recently investigated in Ref.~\cite{bgscat}.
The interaction with the radiation field has been modeled there by a force term entering the equations of motion given by the 4-momentum density of radiation observed in the particle's rest frame with a multiplicative constant factor expressing the strength of the interaction itself.
This approach dates back to the pioneering works of Poynting \cite{Poynting-03} and Robertson \cite{Robertson-37}, who
derived the corrections to the motion of planets in the Solar system due to the scattering of the solar radiation in the context of Newtonian gravity and in the weak field approximation, respectively.
Particles are assumed to interact with the radiation field of an emitting source superimposed on the background by adsorbing and re-emitting radiation, causing a drag force responsible for deviation from geodesic motion, known as the Poynting-Robertson effect. 
The generalization to the framework of general relativity has been developed in Refs. \cite{BiniJS-09,BiniGJSS-11}, where this effect on test particles orbiting in the equatorial plane of a Schwarzschild or Kerr spacetime has been considered, and in Ref.~\cite{vaidyaPR}, where a self-consistent radiation flux was instead used to investigate such a kind of interaction in the Vaidya spherically symmetric spacetime \cite{Vaidya-43}.

In the present paper we consider the scattering of massive particles propagating in the field of a gravitational plane wave and of an electromagnetic wave.
In both cases the wave is sandwiched between two flat Minkowski regions, so that the \lq\lq in'' and \lq\lq out'' regions are unambiguously determined.
The particles will interact differently with the gravitational wave background and the electromagnetic radiation field, so that they will emerge in the outer flat region with different 4-momenta.
The different nature of the host environment will also be evident by comparing the corresponding classical scattering cross sections.

\section{Scattering of particles by a radiation field in a flat spacetime}

Let us consider a Minkowski spacetime with metric written in either Cartesian or related null coordinates as 
\begin{equation}
\label{metricflat2}
ds^2  = -\rmd t^2+\rmd x^2+\rmd y^2+\rmd z^2
= - 2\rmd u \, \rmd v + \rmd x^2 + \rmd y^2\,,
\end{equation}
where
$
u=(t-z)/\sqrt{2}\,,\ v=(t+z)/\sqrt{2}\,.
$ 
The latter form privileges the 3 Killing vectors $\partial_v, \partial_x,\partial_y$ which will remain when a wave zone is introduced later where the metric will depend on $u$.
Figure 1 illustrates the relationships between the coordinates for the case of an interaction strip corresponding to a $u$ coordinate interval $[0,u_1]$.


\begin{figure}[h]
\begin{center}
\includegraphics[scale=1]{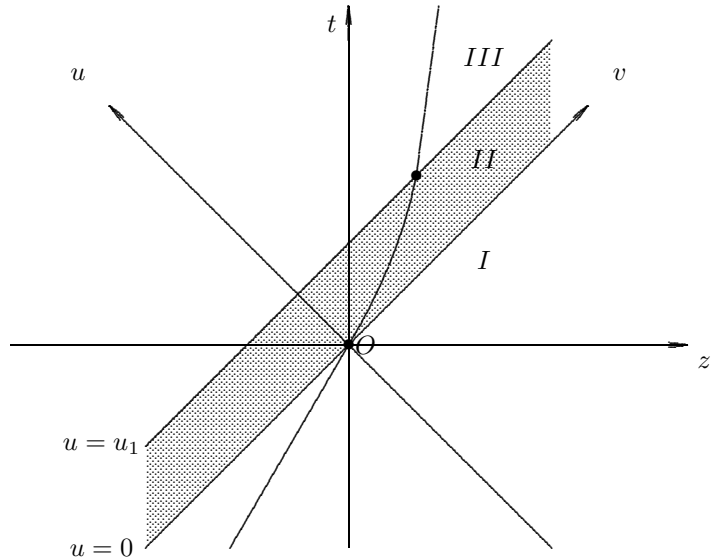}
\end{center}
\caption{The null coordinate relationships in the $t$-$z$ plane (orthogonal to the plane wave fronts aligned with the $x$-$y$ planes) for a sandwich spacetime divided into three zones by the null hypersurfaces $u=0$ and $u=u_1>0$.
Shown also is a suggestive world line of a particle (entering zone II at the origin of coordinates) which is deflected by the radiation field in zone II from its geodesic motion in zones I and III.
}
\label{fig:1}
\end{figure}

It is also useful to introduce a family of \lq\lq static" fiducial observers which are at rest with respect to the spatial coordinates $(x,y,z)$ and characterized by the $4$-velocity vector
$ m=\partial_t$ with the associated adapted
orthonormal spatial triad 
$e_{\hat x}=\partial_x$,
$e_{\hat y}=\partial_y$,
$e_{\hat z}=\partial_z$.

A test particle with rest mass $\mu$ and 4-velocity $U^\alpha=\rmd x^\alpha/\rmd\tau$ (so $U \cdot U=-1$) has 4-momentum 
$P=\mu U$, but we will use the specific 4-momentum, namely the 4-velocity itself: $\tilde P = P/\mu = U$; we drop the tilde notation below and the modifier ``specific." The observer decomposition of $U$ is then (let $a,b,c=1,2,3$)
\beq
\label{repframe}
U = U^\alpha\partial_\alpha
  =\gamma (m+\nu^{\hat a}e_{\hat a})\,,\ 
\gamma =(1-\delta_{\hat a \hat b}\nu^{\hat a}\nu^{\hat b})^{-1/2}\,.
\eeq
For geodesic motion, the constant 4-velocity 
$U=U_{(0)}$
can be parametrized in terms of the conserved specific momenta $p_v$, $p_x$ and $p_y$
(introducing as well $p_\perp^2=p_x^2+p_y^2$) associated with the three Killing vectors mentioned above as
\begin{eqnarray}\label{eq:U0}
U_{(0)} &=& -p_v\left(\partial_u+\frac{1+p_\perp^2}{2p_v^2}\partial_v\right) +p_x\partial_x+p_y\partial_y
\nonumber\\
  &=& -\frac{1}{\sqrt{2}}
 p_v \left(1+\frac{1+p_\perp^2}{2p_v^2}\right)\partial_t+p_x\partial_x + p_y\partial_y
     -\frac{p_v}{\sqrt{2}}\left( -1 +\frac{1+p_\perp^2}{2p_v^2}\right)\partial_z
\,,
\end{eqnarray}
where  $p_v<0$ for $U$ to be future-pointing.
Then the velocity decomposition is
\begin{eqnarray}
\label{compflat}
\gamma_{(0)}&=& -\frac{p_v}{\sqrt{2}}\left(1 +\frac{1+p_\perp^2}{2p_v^2}\right)\,,\nonumber\\
\nu^{\hat x}_{(0)}&=& \frac{p_x}{\gamma_{(0)}}\,,\quad
\nu^{\hat y}_{(0)}= \frac{p_y}{\gamma_{(0)}}\,,\quad
\nu^{\hat z}_{(0)}= -\frac{p_v}{\sqrt{2}\gamma_{(0)}}\left( -1 +\frac{1+p_\perp^2}{2p_v^2}\right)\,,
\end{eqnarray}
which can be easily inverted to yield
\beq
\label{killvsnu0}
p_v=-\frac{\gamma_{(0)}}{\sqrt{2}}(1-\nu^{\hat{z}}_{(0)})\,, \qquad
p_x=\gamma_{(0)}\nu^{\hat{x}}_{(0)}\,, \qquad
p_y=\gamma_{(0)}\nu^{\hat{y}}_{(0)}\,. 
\eeq

Choosing the zero of proper time at the  $u=0$ hyperplane, the corresponding parametric equations of the particle's straight line trajectory are then
\begin{eqnarray}
\label{solregion1}
u &=& -p_v\tau \,,\qquad 
v = \frac{1+p_\perp^2}{2p_v^{2}} u + {v}_0\,, \nonumber\\
x &=& -\frac{p_{x}}{p_v}u + {x}_0\,,\qquad 
y = -\frac{p_{y}}{p_v}u + {y}_0\,,
\end{eqnarray}
so that  $x_0=y_0=z_0=0$ puts the initial position at the origin of coordinates
and
\beq
\label{solregion1b}
t = \frac{1}{\sqrt{2}}\left[\left(\frac{1+p_\perp^{2}}{2p_{v}^{2}}+1\right)u+{v}_0\right]\,,\qquad
z = \frac{1}{\sqrt{2}}\left[\left(\frac{1+p_\perp^{2}}{2p_{v}^{2}}-1\right)u+{v}_0\right]\,.
\eeq

Correspondingly, a photon following a null geodesic path has 4-momentum
\begin{eqnarray}
\label{K_phot}
K&=&-K_v\partial_u- K_u\partial_v+K_x\partial_x+K_y\partial_y\,,
\end{eqnarray}
where  the null condition is $2K_u K_v = K_\perp^2\equiv K_x^2+K_y^2$.
For the special case of photons traveling along the positive $z$-direction, one has $K_x=K_y=K_v=0$ and $K=-K_u\partial_v$, useful for comparison with the nonflat case below.

The observer  decomposition is
\beq
K=\omega_K(m + {\hat \nu}_K)\,, \qquad 
{\hat \nu}_K={\hat \nu}_K^a\partial_a\,,\qquad
{\hat \nu}_K\cdot {\hat \nu}_K=1\,,
\eeq
where 
$\omega_K =K^t=-K_t=-(K_u+K_v)/\sqrt{2}$ 
is the relative energy and the unit vector ${\hat \nu}_K^a={K^a}/{K^t}$ gives the relative direction of propagation with respect to the static observers.

Suppose now that a test radiation field representing a coherent beam of a given frequency fills a certain spacetime region confined to the region between two null hypersurfaces $u=0$ and $u=u_1$ as in Figure 1.
The associated energy-momentum tensor is assumed to be of the form 
\beq
\label{energmomen}
T=\phi_0 K\otimes K\,,
\eeq
where $K$ is the geodesic null vector given by Eq.~(\ref{K_phot}) and $\phi_0$ is a constant representing the associated energy flux.
The geodesic property of $K$ makes $T$ divergence-free, i.e., $\nabla_\alpha T^{\alpha\beta}=0$.

A neutral massive particle moving through the spacetime region occupied by such a radiation field will be scattered in a way which depends on the interaction. 
The simplest way to model this interaction is through the introduction of a \lq\lq radiation force," which is constructed from the energy-momentum tensor introduced in Eq.~(\ref{energmomen}) and is orthogonal to the particle's 4-velocity $U$ (just as the $4$-acceleration vector), so that
\begin{equation}
\label{frad}
f_{\rm (rad)}(U)_\alpha=-\sigma P(U)_{\alpha \beta} T^\beta{}_\mu U^\mu\,,
\end{equation}
where $P(U)=g+U\otimes U$ is the orthogonal projector to $U$ and $\sigma$ models the absorption and re-emission of radiation by the test particle. This force is just proportional to the momentum of the field as observed in the rest frame of the particle.
The equations of motion of the particle thus become
\beq
\label{eqsgen}
\mu a(U)^\alpha=f_{\rm (rad)}(U)^\alpha\,,\qquad 
a(U)^\alpha=\nabla_U U^\alpha\,,
\eeq
or explicitly
\beq
\label{eqsflat}
\frac{\rmd U^\alpha}{\rmd\tau}=-A[K^\alpha +U^\alpha (U\cdot K)](U\cdot K)\,,\qquad 
A=\sigma \phi_0/\mu 
\equiv \tilde\sigma \phi_0
\,.
\eeq
In the case of a particle orbiting a massive source in the presence of a superimposed radiation field, an interaction of this kind leads to a drag force causing deviation from geodesic motion.  This is the so called Poynting-Robertson effect (see Refs.~\cite{Poynting-03,Robertson-37,BiniJS-09,BiniGJSS-11} and references therein). 

The equations of motion (\ref{eqsflat}) then become
\beq\label{eqsflat2}
\frac{\rmd\nu^{\hat a}}{\rmd\tau}=-A \omega_K^2 (\nu^{\hat a}-\nu_K^{\hat a})(1-\nu_K{}_{\hat b}\nu^{\hat b})
\,,
\eeq
whose solution is straightforward assuming $\tau=0$ at the initial null hyperplane $u=0$ 
where $\nu^{\hat a} = \nu_{(0)}^{\hat a}$
\beq
\label{solphot}
\nu^{\hat a}=\nu_K^{\hat a}+\frac{\nu^{\hat a}_{(0)}-\nu_K^{\hat a}}{\left[1+\tau  A\omega_K^2(1-\nu_K{}_{\hat c}\nu^{\hat c}_{(0)})\right]}\,,
\eeq
which can be simplified by introducing the parameter
${1}/{\tau_*}=A\omega_K^2(1-\nu_K{}_{\hat c}\nu^{\hat c}_{(0)})$  to yield
\begin{eqnarray}
\label{nus_eqs}
\nu^{\hat a}=\nu^{\hat a}_K+\frac{\nu^{\hat a}_{(0)}-\nu_K^{\hat a}}{1+\tau/\tau_*}\,, \qquad
\gamma=\gamma_{(0)}\frac{1+\tau/\tau_*}{\sqrt{1  +2\gamma_{(0)}^2 (1-\nu_K{}_{\hat c}\nu^{\hat c}_{(0)}) \tau / \tau_* }}\,.
\end{eqnarray}

The parametric equations for the test particle's trajectory during the interaction with the radiation field are then obtained by integrating the equations $\rmd x^\alpha/\rmd\tau=U^\alpha $, i.e., 
\beq
\frac{\rmd t}{\rmd \tau}=\gamma \,,\quad \frac{\rmd x^{\hat a}}{\rmd \tau}=\gamma \nu^{\hat a}\,.
\eeq
By introducing the notation
\beq
{\mathcal I}(a,b,c,d;\xi)\equiv \int_{0}^\xi \frac{a+b \xi'}{\sqrt{c+d\xi'}}\, 
\rmd \xi'=\frac{2}{3d^2}(3ad-2bc+db\xi') \, \sqrt{c+d\xi'} \bigg\vert_{0}^\xi\,,
\eeq 
the corresponding solution can be explicitly written in the form 
\begin{eqnarray}
\label{orbitaccflat}
t-t_0 &=& \gamma_{(0)}\, {\mathcal I}\left(1,1/\tau_*,1,2\gamma_{(0)}^2 (1-\nu_K{}_{\hat c}\nu^{\hat c}_{(0)})/\tau_*;\tau \right)\,, \nonumber \\
x^a-x^a_0 &=& \gamma_{(0)}\,
{\mathcal I}\left(\nu_{(0)}^{\hat a},\nu^{\hat a}_K/\tau_*,1,2\gamma_{(0)}^2 (1-\nu_K{}_{\hat c}\nu^{\hat c}_{(0)})/\tau_*;\tau \right)\,,
\end{eqnarray}
where the quantities $\gamma_{(0)}$, $\nu^{\hat{a}}_{(0)}$ and $x_0^\alpha$   refer to the (constant) frame components of the particle 4-velocity at the start of the interaction and the initial position there.

In the simplest case of a radiation field composed of photons all propagating along the $z$-direction, i.e., 
with ${\hat \nu}_K^a=\delta^a_z$ and $K=\sqrt{2}\omega_K \partial_v$
the radiation force is 
\beq
\label{frad_flat}
\frac{1}{\mu}f_{\rm (rad)}(U) =-A\, \omega_K^2 \gamma (1-\nu^{\hat z})\left\{ [\gamma^2 (1-\nu^{\hat z})-1]m+\gamma^2 (1-\nu^{\hat z})(\nu^{\hat x}e_{\hat x}+\nu^{\hat y}e_{\hat y})
-[\gamma^2\nu^{\hat z}(\nu^{\hat z}-1)+1]e_{\hat z} \right\}\,,
\eeq
and the general solution (\ref{nus_eqs}) becomes
\beq
\gamma=\frac{\gamma_{(0)}}{\Sigma}[1+A\omega_K^2(1-\nu^{\hat z}_{(0)})\tau ]\,,\qquad
[\nu^{\hat x},\nu^{\hat y},1-\nu^{\hat z}]=\frac{\gamma_{(0)}}{\gamma\Sigma}
\left[\nu^{\hat x}_{(0)},\nu^{\hat y}_{(0)},1-\nu^{\hat z}_{(0)}\right]\,,
\eeq
where $\Sigma=\sqrt{1+2A\omega_K^2\gamma_{(0)}^2(1-\nu^{\hat z}_{(0)})^2\tau }$. 
The parametric equations (\ref{orbitaccflat}) of the accelerated orbit then simplify to
\begin{eqnarray}
t-t_0 &=& \frac{1}{3\gamma_{(0)}(1-\nu^{\hat z}_{(0)})}\left[
\Sigma\tau -\frac{\Sigma-1}{A\omega_K^2\gamma_{(0)}^2(1-\nu^{\hat z}_{(0)})^2}(1-3\gamma_{(0)}^2(1-\nu^{\hat z}_{(0)}))
\right]\,, \nonumber \\
x-x_0 &=& \nu^{\hat x}_{(0)}\frac{\Sigma-1}{A\omega_K^2\gamma_{(0)}(1-\nu^{\hat z}_{(0)})^2}\,, \nonumber \\
y-y_0 &=& \nu^{\hat y}_{(0)}\frac{\Sigma-1}{A\omega_K^2\gamma_{(0)}(1-\nu^{\hat z}_{(0)})^2}\,, \nonumber \\
z-z_0 &=& \frac{1}{3\gamma_{(0)}(1-\nu^{\hat z}_{(0)})}\left[
\Sigma\tau -\frac{\Sigma-1}{A\omega_K^2\gamma_{(0)}^2(1-\nu^{\hat z}_{(0)})^2}(1-3\gamma_{(0)}^2(1-\nu^{\hat z}_{(0)})\nu^{\hat z}_{(0)})
\right] \,.
\end{eqnarray}
The corresponding solutions for $u$ and $v$ are given by
\begin{eqnarray}
u &=& \frac{\Sigma-1}{\sqrt{2}A\omega_K^2\gamma_{(0)}(1-\nu^{\hat z}_{(0)})}\,,\nonumber\\
v - v_0&=& \frac{\sqrt{2}}{3\gamma_{(0)}(1-\nu^{\hat z}_{(0)})}\left[
\Sigma\tau -\frac{\Sigma-1}{A\omega_K^2\gamma_{(0)}^2(1-\nu^{\hat z}_{(0)})^2}\left(1-\frac32\gamma_{(0)}^2(1-\nu^{\hat z}_{(0)}{}^2)\right)
\right]\,.
\end{eqnarray}

The parametric equations for the orbit using $u$ as parameter and re-expressing the initial 3-velocity quantities in terms of the initial momenta are then given by
\begin{eqnarray}
\label{soluvemflat}
u&=& \frac{1}{2p_vA\omega_K^2}\left(1-\sqrt{1+4p_v^2A\omega_K^2\tau }\right)\,,\nonumber\\ 
v - v_0&=& \frac{u}{2p_v^{2}}(1+p_\perp^2)-
\frac{A\omega_K^2u^2}{p_v}\left(1-\frac23p_v A\omega_K^2u\right)\,, \nonumber\\
x - x_0 &=& -\frac{p_{x}}{p_v}u\,,\qquad
y - y_0 = -\frac{p_{y}}{p_v}u\,,
\end{eqnarray}
where 
\beq
\tau_1 = -\frac{u_1}{p_v}(1-p_v A\omega_K^2u)\,
\eeq
relates the proper time interval of the interaction to the interval $u_1$. 
The associated 4-velocity is
\beq
\label{Pemflat}
U=e^{-\zeta_{\rm(tf)}(u)}
\left[-p_v\partial_u-\frac{1}{2p_v}\left(e^{2\zeta_{\rm(tf)}(u)}+p_\perp^2 \right)\partial_v
             +p_x\partial_x+p_y\partial_y\right]\,,
\eeq
where we have introduced the notation (``tf" for test field)
\beq
e^{\zeta_{\rm(tf)}(u)}=1-2p_v A\omega_K^2u\,.
\eeq
This quantity, evaluated at $u=u_1$ ($\tau=\tau_1$) where the interaction with the wave ends, should be compared with the initial 4-velocity  $U_{(0)}$, at $u=0$ ($\tau=0$), given in Eq.~(\ref{eq:U0}). 
In this case, since the spacetime is  flat everywhere, we can image  $U_{(0)}$  (trivially) parallely transported along the particle trajectory up to the same spacetime point where $U$ is located, at the end of the interaction.  The comparison then results  in a 
boost relating these two vectors, namely
\beq
U=\gamma (U,U_{(0)}) \Big[U_{(0)} +||\nu(U,U_{(0)})||\hat \nu(U,U_{(0)})\Big] \,,
\eeq
with the spacelike unit direction-vector of the relative velocity (notation: $U$ with respect to $U_{(0)}$)
given by
\beq
\label{nu_relUU0}
\hat \nu(U,U_{(0)})
=-\frac{P(U_{(0)})\partial_v}{p_v}
=-\frac{1}{p_v}\partial_v -U_{(0)}
=-\frac{1}{\sqrt{2}\omega_Kp_v}K-U_{(0)}\,,
\eeq
where $P(U_{(0)})$ projects orthogonally to $U_{(0)}$ and
$
K=\omega_K (\partial_t+\partial_z)=\sqrt{2}\omega_K \partial_v
$ 
is the photon field.
The relative speed is instead 
\beq
||\nu(U,U_{(0)})||=\tanh (\zeta_{\rm(tf)}(u_1))\,,
\eeq
demonstrating that  $\zeta_{\rm(tf)}(u)$ can be interpreted  as the rapidity boost parameter for the 4-velocity relative to the initial 4-velocity.
Note that this shows that $U$ lies in the plane of $U_{(0)}$ and $K$. In other words the final specific momentum $U$ is just the result of a boost of the initial specific momentum $U_{(0)}$ along the direction of the relative velocity of the wave vector of the radiation field  with respect to it.

This simple analysis can be easily generalized to a sandwich spacetime in which the plane wave zone is a portion of an electrovac  plane-wave spacetime in between two flat spacetime regions as above, either representing the exact gravitational field due to an electromagnetic plane wave or to a gravitational plane wave. The resulting change in 4-momentum or 4-velocity of the test particle from $u=0$ to $u=u_1$ can then be compared with the flat spacetime case with either no interaction or an interaction with a test electromagnetic field as just evaluated. While the scattering by a gravitational plane wave is well known, the electrovac case is not, nor has any comparison been made with the Poynting-Robertson-like interaction, as we will do below.

\section{Scattering of particles by a gravitational plane wave}

Consider the interaction of a test particle with a gravitational radiation field  described by the spacetime metric of an exact gravitational plane wave with a single polarization state ($+$ state) \cite{grif2} traveling in the positive $z$-direction orthogonal to the symmetry planes (with the same relationship between the coordinates as above)
\beq
\label{gw1}
 \rmd s^2 = -\rmd t^2 + F(u)^2\rmd x^2 + G(u)^2\rmd y^2+ \rmd z^2
= - 2\rmd u \rmd v +F(u)^2\rmd x^2+G(u)^2\rmd y^2\,,
\eeq
with 
\beq
\label{FGgw}
F(u)=\cos(b_{\rm (gw)}u)\,, \qquad
G(u)=\cosh(b_{\rm (gw)}u)\,,
\eeq
where $\omega_{\rm (gw)}=b_{\rm (gw)}/\sqrt{2}$ is the frequency
of the gravitational wave under consideration, and $s=b_{\rm (gw)} u = \omega_{\rm (gw)}(t-z)$ is a convenient combination used below. We continue to use the same static frame as in the flat case.

The gravitational wave is sandwiched between two Minkowskian regions $u\in (-\infty, 0)\cup (u_1, \infty)$, and the metric would have a coordinate horizon at $b_{\rm (gw)}u=\frac{\pi}{2}$ where the metric is degenerate but this is avoided by restricting the coordinate $u$ to the interval $[0,u_1]$ with $b_{\rm (gw)}u_1<\frac{\pi}{2}$. 
The matching conditions impose restrictions on the metric functions $F$ and $G$ before and after the passage of the wave where the spacetime is Minkowskian.
As discussed in detail by Rindler in Ref.~\cite{rindler} (see this reference for a more detailed account of exact plane gravitational waves),
a possible choice  to extend the metric for all values of $u$  is the following
\beq\label{Eq8}
\begin{array}{c|c|c}
& F & G\\
\hline 
\hbox{\rm(I) }  & 1  & 1\\
\hbox{\rm(II) } & \cos(b_{\rm (gw)}u) & \cosh(b_{\rm (gw)}u)\\
\hbox{\rm(III) } &(\alpha+\beta u) & (\gamma+\delta u)
\end{array}
\eeq
where labels I, II and III refer to the in-zone ($u\le 0$), the wave-zone ($0<u<u_1$) and the out-zone $(u\ge u_1)$, respectively.
Values of the constants $\alpha$, $\beta$, $\gamma$ and $\delta$ can be completely determined by requiring $C^1$ regularity conditions at the boundaries $u=0$ and $u=u_1$ of the sandwich, that is
\beq\label{regconds}
\begin{array}{lll}
&F(0)=1=G(0)\,,\qquad
&F'(0)=0=G'(0)\,, \\
&F(u_1)=\alpha+\beta u_1\,,  
&G(u_1)=\gamma+\delta u_1\,, \\
&F'(u_1)=\beta\,,  
&G'(u_1)=\delta\,,
\end{array}
\eeq
which in this case imply
\beq
\label{costanti}
\begin{array}{lllll}
\alpha &=&\cos(b_{\rm (gw)}u_1) + b_{\rm (gw)}u_1\sin(b_{\rm (gw)}u_1)\,,\qquad \,\,\,\,\,\,\beta &=&-b_{\rm (gw)}\sin(b_{\rm (gw)}u_1) \,,\cr
\gamma &=& \cosh(b_{\rm (gw)}u_1) - b_{\rm (gw)}u_1\sinh(b_{\rm (gw)}u_1)\,,\qquad  \delta &=&b_{\rm (gw)}\sinh(b_{\rm (gw)}u_1)\,.
\end{array}
\eeq

Let us consider the wave region (II), with functions $F$ and $G$ given by Eq.~(\ref{FGgw}). 
As in the flat spacetime case, a test particle with mass $\mu$ entering the wave region follows a geodesic path with 4-velocity $U$ and associated 4-momentum $P=\mu U$  given by (see, e.g., Ref.~\cite{bp})
\beq
\label{Pgeogw}
U=-p_v\partial_u-\frac{1}{2p_v}\left(1+\frac{p_x^2}{F(u)^2}+\frac{p_y^2}{G(u)^2}\right)\partial_v+\frac{p_x}{F(u)^2}\partial_x+\frac{p_y}{G(u)^2}\partial_y\,,
\eeq
where the conserved specific momenta $p_v$, $p_x$ and $p_y$ still allow the complete integration of the geodesic equations.
Using the explicit form of the metric functions $F$ and $G$ valid in the wave-zone and imposing the matching at the boundary I--II where the geodesics join at the spacetime point with coordinates $(0,v_0,x_0,y_0)$ then gives
\begin{eqnarray}
\label{solregion2gwexplfin}
u &=& -  p_v \tau \,,\quad 
v = \frac{1}{2  p_v^{2}}\left(u+\frac{  p_x^2}{b_{\rm (gw)}}\tan(b_{\rm (gw)}u)+\frac{  p_y^2}{b_{\rm (gw)}}\tanh(b_{\rm (gw)}u)\right) + {v}_{0}\,, \nonumber\\
x &=& -\frac{  p_{x}}{b_{\rm (gw)}  p_v}\tan(b_{\rm (gw)}u) + {x}_{0}\,,\quad 
y = -\frac{  p_{y}}{b_{\rm (gw)}  p_v}\tanh(b_{\rm (gw)}u) + {y}_{0}\,.
\end{eqnarray}
Clearly, these geodesic world lines should be matched with the straight lines of the in-zone at $u=0$.

The geodesic 4-velocity in the inertial coordinates and with the metric functions conveniently re-expressed in terms of $s=\omega_{\rm (gw)} (t-z)$ is
\begin{eqnarray}
\label{Pgeogwcart}
U&=&-\frac{p_v}{\sqrt{2}}
\left[ 1 + \frac{1}{2p_v^2}\left(1+\frac{p_x^2}{\cos^2 s}+\frac{p_y^2}{\cosh^2 s}\right)\right]\partial_t
+\frac{p_x}{\cos^2 s}\partial_x+\frac{p_y}{\cosh^2 s}\partial_y\nonumber\\
&&-\frac{p_v}{\sqrt{2}}
\left[ -1 + \frac{1}{2p_v^2}\left(1+\frac{p_x^2}{\cos^2 s }+\frac{p_y^2}{\cosh^2 s }\right)\right]\partial_z\,.
\end{eqnarray}
Coordinate and frame components of the 4-velocity are now related by 
\beq
\label{Ucompts}
U^t=\gamma\,,\quad
\frac{U^x}{U^t }=\frac{\nu^{\hat x}}{\cos s} \,,\quad
\frac{U^y}{U^t }=\frac{\nu^{\hat y}}{\cosh s} \,,\quad
\frac{U^z}{U^t }= \nu^{\hat z}\,. 
\eeq
Using the relations Eq.~(\ref{killvsnu0}) at $s=0$ to express the Killing constants $(  p_v,  p_x,  p_y)$ in terms of the initial values $\nu^{\hat{a}}_{(0)}\equiv\nu^{\hat{a}}(0)$ at the start of the interaction, one finds with some manipulation
\begin{eqnarray}
\gamma&=&\frac{\gamma_{(0)}}{2(1-\nu^{\hat{z}}_{(0)})}\frac{V(s)}{\cos^2s\cosh^2s}\,, \nonumber\\
{}[\nu^{\hat{x}},\nu^{\hat{y}},1-\nu^{\hat{z}}]
&=&2(1-\nu^{\hat{z}}_{(0)})\frac{\cos s\cosh s}{V(s)}
\left[\nu^{\hat{x}}_{(0)}\cosh s,\nu^{\hat{y}}_{(0)}\cos s,(1-\nu^{\hat{z}}_{(0)})\cos s\cosh s\right]\,,
\end{eqnarray}
where
\beq
V(s)=\left[2(1-\nu^{\hat{z}}_{(0)})\cos^2s+\nu^{\hat{x}}_{(0)}{}^2\sin^2s\right]\cosh^2s-\nu^{\hat{y}}_{(0)}{}^2\cos^2s\sinh^2s\,.
\eeq
Finally, the parametric equations for the particle's geodesic orbit are
\begin{eqnarray}
\label{trajGW}
t -t_0&=& \frac{1}{\omega_{\rm (gw)}(1-\nu^{\hat{z}}_{(0)})}\left[\left(1-\frac{\nu^{\hat{x}}_{(0)}{}^2+\nu^{\hat{y}}_{(0)}{}^2}{2(1-\nu^{\hat{z}}_{(0)})}\right)s+\frac{\nu^{\hat{x}}_{(0)}{}^2\tan s+\nu^{\hat{y}}_{(0)}{}^2\tanh s}{2(1-\nu^{\hat{z}}_{(0)})}\right]\,, \nonumber \\
z-z_0 &=& \frac{1}{\omega_{\rm (gw)}(1-\nu^{\hat{z}}_{(0)})}\left[\left(\nu^{\hat{z}}_{(0)}-\frac{\nu^{\hat{x}}_{(0)}{}^2+\nu^{\hat{y}}_{(0)}{}^2}{2(1-\nu^{\hat{z}}_{(0)})}\right)s+\frac{\nu^{\hat{x}}_{(0)}{}^2\tan s+\nu^{\hat{y}}_{(0)}{}^2\tanh s}{2(1-\nu^{\hat{z}}_{(0)})}\right]\,,
\nonumber \\
x-x_0 &=& \frac{\nu^{\hat{x}}_{(0)}\tan s}{\omega_{\rm (gw)}(1-\nu^{\hat{z}}_{(0)})}\,,\quad 
y-y_0 = \frac{\nu^{\hat{y}}_{(0)}\tanh s}{\omega_{\rm (gw)}(1-\nu^{\hat{z}}_{(0)})}\,,\quad 
s = \omega_{\rm (gw)}\gamma_{(0)}(1-\nu^{\hat{z}}_{(0)})\tau\,,
\end{eqnarray}
where $(t_0,x_0,y_0,z_0)$ denote the coordinates of the spacetime point where the interaction between the test particle and the gravitational wave starts.

\section{Scattering of particles by an electromagnetic plane wave}

Now instead let the test particle interact with a photon radiation field in the gravitational field generated by an electromagnetic plane wave propagating along the positive $z$-axis exactly as in the flat spacetime case in Sec. II.
The corresponding conformally flat line element found by Griffiths \cite{grif} is given by Eq.~(\ref{gw1}) with functions 
\beq
\label{FGem}
F=\cos(b_{\rm (em)}u)=G\,,
\eeq
differing from the corresponding gravitational wave case only by a trigonometric rather than hyperbolic cosine appearing in $G$, so that the above analysis with the additional interaction with the radiation field is easily repeated as done in Ref.~\cite{bgscat}, allowing a comparison between these two cases as well as with the flat one.
However,
the present case corresponds to a nonvacuum spacetime which is a solution of the Einstein equations with energy-momentum tensor
\beq
\label{Tmunu}
T=\phi_0 K \otimes K\,,\quad  K = {b_{\rm (em)}} \partial_v =\sqrt{2} \omega_{\rm (em)} \partial_v \,,
\eeq
where  $\phi_0=1/4\pi$ and $\omega_{\rm (em)}$ is the frequency of the wave.
This corresponds to the flat case of Sec. II with $K_u<0$ and $K_x=K_y=K_v=0$ and $ \omega_{\rm (em)}=\omega_K$, which makes the energy-momentum tensors agree.
For convenience we introduce the parameter 
$s=b_{\rm (em)} u = \omega_{\rm (em)}(t-z)$.

As in the previous section, the metric would have a coordinate horizon at $ b_{\rm (em)}u=\pi/2$ but this is avoided by restricting the coordinate $u$ to the interval $[0,u_1]$ with $u_1<\pi/(2b_{\rm (em)})$.
Similarly 
let the electromagnetic wave spacetime be sandwiched between two Minkowskian regions $u\in (-\infty, 0)\cup (u_1, \infty)$, again as in Figure 1. 
The matching conditions (\ref{regconds}) at the two null hypersurface boundaries now imply 
\beq
\label{costanti2}
\alpha =\cos(b_{\rm (em)}u_1) + b_{\rm (em)}u_1\sin(b_{\rm (em)}u_1)=\gamma\,,\qquad
\beta=-b_{\rm (em)}\sin(b_{\rm (em)}u_1)=\delta\,.
\eeq

Again consider the behavior of neutral test particles in such a spacetime with the additional interaction with the radiation field deflecting them from geodesic motion. However,
now the radiation field is not a test field superimposed on a given gravitational background, so that the treatment is self-consistent.

The observer decomposition of the radiation force of Eqs.~(\ref{frad}) and (\ref{eqsgen}) is formally the same as in Eq. (\ref{frad_flat}), with $\omega_K$ replaced by $\omega_{\rm (em)}$ and the parameter $A$ defined as in Eq. (\ref{eqsflat}).
The flat spacetime equations of motion (\ref{eqsflat2}) with $\nu_K^{\hat a}= \delta^{\hat a}_{z}$ 
acquire an extra term proportional to $\omega_{\rm (em)}$ which now explicitly depends on the coordinate $u$ through  $s$
\begin{eqnarray}
\label{eqmoto}
\frac{\rmd \nu^{\hat x}}{\rmd \tau} &=& - A\, \omega_{\rm (em)}^2(1-\nu^{\hat z})\nu^{\hat x} -\omega_{\rm (em)} \gamma \nu^{\hat x}\tan s \,(\nu^{\hat x}{}^2+\nu^{\hat y}{}^2+\nu^{\hat z}-1)\,,\nonumber\\
\frac{\rmd \nu^{\hat y}}{\rmd \tau} &=& - A\, \omega_{\rm (em)}^2 (1-\nu^{\hat z})\nu^{\hat y}-\omega_{\rm (em)}\gamma \nu^{\hat y}\tan s \,(\nu^{\hat x}{}^2+\nu^{\hat y}{}^2+\nu^{\hat z}-1)\,,\nonumber\\
\frac{\rmd \nu^{\hat z}}{\rmd \tau} &=&  A\, \omega_{\rm (em)}^2 (1-\nu^{\hat z})^2+\omega_{\rm (em)}\gamma\tan s \,(\nu^{\hat y}{}^2+\nu^{\hat x}{}^2) (1-\nu^{\hat z})\,.
\end{eqnarray}
These must be completed with the evolution equations for $t$, $x$, $y$ and $z$ (see Eq.~(\ref{Ucompts})), i.e.,
\beq
\label{eqevol}
\frac{\rmd t}{\rmd \tau}=\gamma\,,\qquad 
\frac{\rmd x}{\rmd \tau}=\frac{\gamma\nu^{\hat x}}{\cos s}\,,\qquad 
\frac{\rmd y}{\rmd \tau}=\frac{\gamma\nu^{\hat y}}{\cos s}\,,\qquad 
\frac{\rmd z}{\rmd \tau}=\gamma \nu^{\hat z}\,,
\eeq
which can be integrated exactly, first re-expressing the derivatives in terms of $s$ through
$\rmd s/\rmd\tau=\omega_{\rm (em)} \gamma (1-\nu^{\hat z}) $.
This simplifies the velocity equations to
\begin{eqnarray}
\frac{\rmd \nu^{\hat x}}{\rmd s} 
&=& -\frac{A}{\gamma} \omega_{\rm (em)} \nu^{\hat x} - \frac{\nu^{\hat x}}{1-\nu^{\hat z}} (\nu^{\hat x}{}^2+\nu^{\hat y}{}^2+\nu^{\hat z}-1)\tan s\,,\nonumber\\
\frac{\rmd \nu^{\hat y}}{\rmd s} 
&=& -\frac{A}{\gamma} \omega_{\rm (em)} \nu^{\hat y}- \frac{ \nu^{\hat y}}{1-\nu^{\hat z}}(\nu^{\hat x}{}^2+\nu^{\hat y}{}^2+\nu^{\hat z}-1)\tan s\,,\nonumber\\
\frac{\rmd \nu^{\hat z}}{\rmd s} 
&=& \frac{A}{\gamma} \omega_{\rm (em)}  (1-\nu^{\hat z}) +   (\nu^{\hat y}{}^2+\nu^{\hat x}{}^2)  \tan s\,.
\end{eqnarray}
The corresponding solutions are then easily obtained
\begin{eqnarray}
\label{nusolfin}
\gamma&=&\frac{\gamma_{(0)}}{\cos^2s}\frac{1+(1-\nu^{\hat z}_{(0)})(W(s)\cos^2s-W(0))}{1+A\gamma_{(0)}(1-\nu^{\hat z}_{(0)})  \omega_{\rm (em)} s}\,, \nonumber\\
{}[\nu^{\hat x},\nu^{\hat y}, 1-\nu^{\hat z}]&=&\frac{\cos s}{1+(1-\nu^{\hat z}_{(0)})(W(s)\cos^2s-W(0))}[\nu^{\hat x}_{(0)},\nu^{\hat y}_{(0)}, (1-\nu^{\hat z}_{(0)}) \cos s ]\,,
\end{eqnarray}
where $\nu^{\hat{a}}_{(0)}\equiv\nu^{\hat{a}}(0)$ and 
\beq
W(s)=\frac{1}{2}+\frac{1}{2}\frac{(1+A\gamma_{(0)}(1-\nu^{\hat z}_{(0)})  \omega_{\rm (em)} s)^2}{\gamma_{(0)}^2(1-\nu^{\hat z}_{(0)})^2}\,.
\eeq
When $ A=0$ (geodesic case) the solution is still given by Eq.~(\ref{nusolfin}) with $W(s)=W(0)$.
As in the previous section
the integration of the equations of motion has been carried out by assuming  that the interaction starts at a proper time $\tau=0$ associated with $s=0$, and that before the interaction the test particle moves along geodesic lines described by Eqs.~(\ref{solregion1}) and (\ref{solregion1b}). Again the values $\nu^{\hat{a}}_{(0)}$ refer to the particle's initial spatial velocity at the start of the interaction, whose relation with the Killing constants $(  p_v,  p_x,  p_y)$ is still given by Eq.~(\ref{killvsnu0}).

Using Eqs. (\ref{nusolfin}), the equation for $s$ then becomes 
\beq
\label{eqsditauem2}
\frac{\rmd s}{\rmd \tau}=\frac{\gamma_{(0)} (1-\nu^{\hat z}_{(0)})\omega_{\rm (em)}}{1+A\gamma_{(0)} (1-\nu^{\hat z}_{(0)}) \omega_{\rm (em)} s}\,,
\eeq
whose solution is 
\beq
\label{sditausol}
s=\frac{\sqrt{1+2A\omega_{\rm (em)}^2\gamma_{(0)}^2 (1-\nu^{\hat z}_{(0)})^2 \tau }-1}{A\omega_{\rm (em)}\gamma_{(0)} (1-\nu^{\hat z}_{(0)}) }\,.
\eeq

Eqs.~(\ref{eqevol}) can then be integrated to obtain the solution for the accelerated orbit (see Ref.~\cite{bgscat} for details)
leading finally to
the parametric equations for the orbit in terms of the coordinates $(u,v,y,z)$ with $u$ as the parameter
\begin{eqnarray}
\label{solregion2uvem}
u  &=& \frac{s }{b_{\rm (em)}}=\frac{1}{A  p_vb_{\rm (em)}^2 }\left(1-\sqrt{1+2A  p_v^2b_{\rm (em)}^2 \tau }\right)\,,\nonumber\\ 
v - v_0&=& \frac{1}{2  p_v^{2}}\left(u+\frac{  p_\perp^2}{b_{\rm (em)}}\tan(b_{\rm (em)}u)\right)-
\frac{ A b_{\rm (em)}^2u^2}{ 2 p_v}\left(1-\frac{A}3  p_v b_{\rm (em)}^2u\right)\,, \nonumber\\
x  - x_0 &=& -\frac{  p_{x}}{b_{\rm (em)}  p_v}\tan(b_{\rm (em)}u)\,,\qquad
y  - y_0 = -\frac{  p_{y}}{b_{\rm (em)}  p_v}\tan(b_{\rm (em)}u)\,,
\end{eqnarray}
with associated 4-velocity
\beq
\label{Pem}
U=e^{-\zeta_{\rm (em)}(u)}
\left\{-p_v\partial_u-\frac{1}{2p_v}\left[e^{2\zeta_{\rm (em)}(u)}+\frac{p_\perp^2}{\cos^2(b_{\rm (em)}u)}\right]\partial_v
+\frac{p_x}{\cos^2(b_{\rm (em)}u)}\partial_x+\frac{p_y}{\cos^2(b_{\rm (em)}u)}\partial_y\right\}\,,
\eeq
where we have introduced the notation
\beq
e^{\zeta_{\rm (em)}(u)} = 1-A  p_v b_{\rm (em)}^2u\,.
\eeq

\section{Test particle motion after the interaction with a radiation field}

Let us now consider a test particle emerging from its interaction in region II with a yet unspecified radiation field (including the flat case with a test radiation field)
entering the flat spacetime region  III (see Eq.~(\ref{Eq8})) at the point $P_1$ with coordinates $(u_1,v_1,x_1,y_1)$ associated with a proper time value $\tau_1$.
Although the spacetime in region III is flat, the metric functions $F(u)$ and $G(u)$ for both the case of electromagnetic and gravitational wave do not have the value 1 associated with flat coordinates.
In fact, they can be represented by 
\beq
F(u)=\alpha+\beta u\,,\qquad
G(u)=\gamma+\delta u\,.
\eeq 
Clearly, this representation also holds in the flat case with $\alpha=1=\gamma$ and $\beta=0=\delta$.
Standard Cartesian coordinates must be obtained by two successive coordinate transformations, namely
$(u,v,x,y) \rightarrow ({\mathcal U},V,X,Y)$
\begin{eqnarray}
\label{coordtrans}
{\mathcal U}=u\,,\qquad
X =F(u)\, x \,,\qquad
Y =G(u)\, y \,,\qquad 
V=v+\frac12F(u)F'(u)\, x^{2}+\frac12G(u)G'(u)\, y^{2}\,,
\end{eqnarray}
for which $\partial_V = \partial_v$ and then
 $({\mathcal U},V,X,Y) \rightarrow (T,X,Y,Z)$
\begin{eqnarray}
T=\frac{{\mathcal U}+V}{\sqrt{2}}\,,\qquad Z=\frac{V-{\mathcal U}}{\sqrt{2}}\,,\qquad  X=X\,,\qquad Y=Y\,.
\end{eqnarray}

Let us denote the specific 4-momentum in region III and in $({\mathcal U},V,X,Y)$ coordinates by 
\beq
\label{Uregion3}
U = -Q_V\left(\partial_{\mathcal U}+\frac{1+Q_\perp^2}{2Q_V^2}\partial_V\right) +Q_X\partial_X+Q_Y\partial_Y\,,
\eeq
where $Q_V,Q_X,Q_Y$ are constant.
The emerging particle 4-velocity and the parametric equations for its trajectory are then explicitly obtained (in both coordinate systems) by imposing matching conditions at the boundary II--III where $\tau = \tau_1$, and will be discussed below in the three different cases. 

Finally consider a collection of particles labeled by their initial coordinates $x_0$ and $y_0$ along the transverse directions $x$ and $y$ to the wave propagation.
Particles scattered by the wave pulse will have different outgoing momentum 4-vectors, depending on their initial data.
The matching at the boundary II--III of the wave-zone and out-zone 4-momenta provide a map between the transverse components of the 4-momentum in any spacelike plane
associated with the static observer's rest space in the final Minkowski region and the initial location of those particles in a similar plane in the initial Minkowski region.
Therefore, one can define a classical differential scattering cross section associated with this transverse scattering map in terms of the outgoing momentum components as follows \cite{garriga}
\beq
\label{sigmadef}
\rmd\sigma_{\rm class}=\rmd x_0\rmd y_0=|J|\rmd Q_{X}\rmd Q_{Y}\,,
\eeq
where $J$ denotes the Jacobian of the transformation between $(Q_{X},Q_{Y})$ and $(x_0,y_0)$.

\subsection{Flat spacetime with test radiation field}

In the simplest case of a test radiation field  superimposed on a flat spacetime we find (see Eq.~(\ref{Pemflat}))
\beq
Q_V=p_ve^{-\zeta_{\rm(tf)}(u_1)}\,,\qquad 
Q_X=p_x e^{-\zeta_{\rm(tf)}(u_1)}\,,\qquad 
Q_Y=p_y e^{-\zeta_{\rm(tf)}(u_1)}\,.
\eeq
Thus the transverse differential scattering cross section vanishes in this case. The effect of the test field on the particle's 4-velocity has been examined in Sec. II, considering the initial and final $4$-velocity vectors, as given by Eqs. (\ref{soluvemflat}) and (\ref{Pemflat}), in flat spacetime. In the case of a test radiation field the vectors are related by a boost
\beq
\label{UminusU0tf}
U-U_{(0)} \equiv \Delta U_{\rm (tf)}=\left(e^{-\zeta_{\rm(tf)}(u_1)}-1\right)\left[
-p_v\partial_u-\frac{1}{2p_v}\left(-e^{\zeta_{\rm(tf)}(u_1)}+p_\perp^2\right)\partial_v+p_x\partial_x+p_y\partial_y
\right]\,.
\eeq
The effect of the wave on the particle's $4$-velocity can be also summarized by a boost if one considers both the initial and the final $4$-velocity vectors in the same flat spacetime. We can write
\beq
U=\gamma(U,U_{(0)})[U_{(0)}+\nu(U,U_{(0)})]
\eeq
with
\beq
\label{gamma_nu}
\gamma(U,U_{(0)})=1-U_{(0)}\cdot \Delta U_{\rm (tf)}\,,\qquad \nu(U,U_{(0)})=\frac{P(U_{(0)})\Delta U_{\rm (tf)}}{1-U_{(0)}\cdot \Delta U_{\rm (tf)}}\,,
\eeq
where $P(U_{(0)})$ projects orthogonally to $U_{(0)}$ and the scalar product here refers to the flat spacetime metric.

\subsection{Gravitational wave radiation field}

In order to obtain the values of the constant components of the emerging 4-momentum we first apply the coordinate tranformation (\ref{coordtrans}) to the 4-velocity (\ref{Uregion3}).
Next we require the latter to match at the boundary II--III where $\tau = \tau_1=-u_1/p_v$, i.e., at the spacetime point $P_1$ with coordinates $(u_1,v_1,x_1,y_1)$, with the wave-zone 4-velocity (\ref{Pgeogw}) with functions $F$ and $G$ given by Eq.~(\ref{FGgw}).
By identifying the components there we finally get the result
\beq
\label{IIImomenta2}
Q_V = p_v\,,\quad 
Q_X = p_v \sin(b_{\rm (gw)} u_1)b_{\rm (gw)} x_0 +p_x \cos(b_{\rm (gw)} u_1)\,,\quad 
Q_Y = -p_v \sinh(b_{\rm (gw)} u_1)b_{\rm (gw)} y_0 +p_y \cosh(b_{\rm (gw)} u_1)\,,
\eeq
where the following relations have been used 
\begin{eqnarray}
\label{match23}
v_{1} &=& \frac{1}{2  p_v^2}\left[u_1 +\frac{  p_x^2}{b_{\rm (gw)}}\tan\left(b_{\rm (gw)} u_1\right)+\frac{  p_y^2}{b_{\rm (gw)}}\tanh\left(b_{\rm (gw)} u_1\right)\right]+v_0\,,\nonumber\\
x_{1} &=& -\frac{  p_{x}}{b_{\rm (gw)}  p_v}\tan\left(b_{\rm (gw)} u_1\right)+x_{0}\,, \quad 
y_{1} = -\frac{  p_{y}}{b_{\rm (gw)}  p_v}\tanh\left(b_{\rm (gw)} u_1\right)+y_{0}\,,
\end{eqnarray}
to re-express the coordinates at the boundary $P_1$ in terms of those of $P_0$ associated with $\tau=0$, where the interaction between the test particle and the gravitational wave starts.
Note that also the flat spacetime coordinate frames $\{\partial_{\mathcal U},\partial_V,\partial_X,\partial_Y\}$ and $\{\partial_u,\partial_v,\partial_x,\partial_y\}$ have been identified to make the comparison, and that the momentum $p_v$ is conserved here.

The differential (transverse) scattering cross section (\ref{sigmadef}) is then given by
\beq
\rmd\sigma_{\rm class}^{\rm(gw)}=\frac{\rmd Q_{X}\rmd Q_{Y}}{p_v^2b_{\rm (gw)}^2\sin(b_{\rm (gw)} u_1)\sinh(b_{\rm (gw)} u_1)}\,.
\eeq
The effect of the wave on the particle's $4$-velocity can be also summarized by a boost if one considers both the initial and the final $4$-velocity vectors in the same flat spacetime. In this sense, by using Eq. (\ref{Uregion3}) and its analogous for $U_{(0)}$ before the passage of the wave (i.e., with $Q_V,Q_X,Q_Y$ replaced by $p_v,p_x,p_y$), we can write
\beq
\label{UU0GW}
U-U_{(0)}\equiv \Delta U_{\rm (gw)}=-\frac{1}{2p_v}(Q_\perp^2-p_\perp^2)\partial_v+(Q_X-p_x)\partial_x+(Q_Y-p_y)\partial_y\,,
\eeq
as in Eq. (\ref{UminusU0tf}), so that the relative decomposition $U=\gamma(U,U_{(0)})(U_{(0)}+\nu(U,U_{(0)})$ is accomplished with 
\beq
\label{gamma_nu_gw}
\gamma(U,U_{(0)})=1-U_{(0)}\cdot \Delta U_{\rm (gw)}\,,\qquad \nu(U,U_{(0)})=\frac{P(U_{(0)})\Delta U_{\rm (gw)}}{1-U_{(0)}\cdot \Delta U_{\rm (gw)}}\,,
\eeq
as in Eq. (\ref{gamma_nu}), where $P(U_{(0)})$ projects orthogonally to $U_{(0)}$ and the scalar product here refers to the flat spacetime metric. The direct evaluation of the relative velocity $\nu(U,U_{(0)})$ follows straightforwardly from Eq. (\ref{IIImomenta2}). Note that Eq. (\ref{UU0GW}) is the curved spacetime counterpart of Eq. (\ref{nu_relUU0}) of the flat case examined previously. For instance, assuming $x_0=0=y_0$, from Eqs. (\ref{IIImomenta2}) we have
\beq
Q_X=p_x \cos(b_{\rm (gw)} u_1)\,,\quad
Q_Y=p_y \cosh(b_{\rm (gw)} u_1)\,,
\eeq
and hence $\Delta U_{\rm (gw)}$ becomes
\beq
\Delta U_{\rm (gw)}=-\frac{1}{2p_v}[-p_x^2 \sin^2(b_{\rm (gw)} u_1)+p_y^2\sinh^2(b_{\rm (gw)} u_1)]\partial_v+p_x[\cos(b_{\rm (gw)} u_1)-1]\partial_x+p_y[\cosh(b_{\rm (gw)} u_1)-1]\partial_y\,.
\eeq

\subsection{Electromagnetic wave radiation field}

In the case of the spacetime of an electromagnetic wave, the matching conditions at $P_1$ with coordinates $(u_1,v_1,x_1,y_1)$ give the following value of the proper time
\beq
\tau_1=-\frac{u_1}{  p_v}\left(1- \frac12 A p_v b_{\rm (em)}^2 u_1\right)\,.
\eeq
The relation between \lq\lq in'' and \lq\lq out'' momenta in this case is 
\begin{eqnarray}
\label{IIImomenta_em}
Q_V &=& p_ve^{-\zeta_{\rm (em)}(u_1)}\,,\nonumber\\ 
Q_X &=& [p_v \sin(b_{\rm (em)} u_1)b_{\rm (em)} x_0 +p_x \cos(b_{\rm (em)} u_1)]e^{-\zeta_{\rm (em)}(u_1)}\,,\nonumber\\ 
Q_Y &=& [p_v \sin(b_{\rm (em)} u_1)b_{\rm (em)} y_0 +p_y \cos(b_{\rm (em)} u_1)]e^{-\zeta_{\rm (em)}(u_1)}\,,
\end{eqnarray}
where the following relations have been used 
\begin{eqnarray}
\label{match23em}
v_{1} &=& \frac{1}{2  p_v^2}\left[u_1 +\frac{  p_\perp^2}{b_{\rm (em)}}\tan\left(b_{\rm (em)} u_1\right)\right]-
\frac{ \frac12 A b_{\rm (em)}^2u_1^2}{  p_v}\left(1-\frac13  A p_v b_{\rm (em)}^2u_1\right)+v_0\,,\nonumber\\
x_{1} &=& -\frac{  p_{x}}{b_{\rm (em)}  p_v}\tan\left(b_{\rm (em)} u_1\right)+x_{0}\,, \quad 
y_{1} = -\frac{  p_{y}}{b_{\rm (em)}  p_v}\tan\left(b_{\rm (em)} u_1\right)+y_{0}\,.
\end{eqnarray}

The differential scattering cross section (\ref{sigmadef}) is then given by
\beq
\rmd\sigma_{\rm class}^{\rm(em)}=\frac{e^{2\zeta_{\rm (em)}(u_1)}\rmd Q_{X}\rmd Q_{Y}}{p_v^2b_{\rm (em)}^2\sin^2(b_{\rm (em)} u_1)}\,.
\eeq
The effect of the wave on the particle's $4$-velocity can be similarly summarized by a boost if one considers both the initial and the final $4$-velocity vectors in the same flat spacetime, identifying $\{\partial_{\mathcal U},\partial_V,\partial_X,\partial_Y\}$ with $\{\partial_u,\partial_v,\partial_x,\partial_y\}$ in order to make the comparison. 
Expressing $U$ in region III the same form as Eq. (\ref{Uregion3}) (but now taking into account Eq. (\ref{IIImomenta_em})) and comparing it with the original $U_{(0)}$ before the passage of the electromagnetic wave, we can now write the final difference as
$U-U_{(0)}\equiv \Delta U_{\rm (em)}$ with 
\beq
\label{UU0EM}
\Delta U_{\rm (em)}=-(Q_V-p_v)\partial_u+\frac{1}{2Q_Vp_v}[(1+p_\perp^2)Q_V-(1+Q_\perp^2)p_v]\partial_v+(Q_X-p_x)\partial_x+(Q_Y-p_y)\partial_y\,.
\eeq
Similarly, the relative decomposition $U=\gamma(U,U_{(0)})(U_{(0)}+\nu(U,U_{(0)}))$ is accomplished with the equivalent of Eq. (\ref{gamma_nu_gw}).
Here, the direct evaluation of the relative velocity $\nu(U,U_{(0)})$ follows straightforwardly from Eq. (\ref{IIImomenta_em}). 
For instance, for $x_0=0=y_0$ the difference $\Delta U_{\rm (em)}$ becomes
\begin{eqnarray}
\Delta U_{\rm (em)}&=&-p_v\left(e^{-\zeta_{\rm (em)}(u_1)}-1\right)\partial_u
-\frac{1}{2p_v}\left[\left(e^{\zeta_{\rm (em)}(u_1)}-1\right)+p_\perp^2\left(\cos^2(b_{\rm (em)}u)e^{-\zeta_{\rm (em)}(u_1)}-1\right)\right]\partial_v\nonumber\\
&&+\left(e^{-\zeta_{\rm (em)}(u_1)}\cos(b_{\rm (em)} u_1)-1\right)(p_x\partial_x+p_y\partial_y)\,.
\end{eqnarray}

\section{Discussion}

Let a massive test particle  be scattered by a radiation field filling a spacetime region and imagine that the source of radiation is unknown.
For the purpose of the present investigation we have considered three different kinds of radiation: a photon test field in a flat spacetime background, an exact solution of the Einstein field equations for a strong plane gravitational wave (with single polarization state for simplicity), an exact solution of the Einstein-Maxwell equations representing the curved spacetime associated with a plane electromagnetic wave.
The effect of the interaction in all cases is a change in the linear momentum of the particle from its initial state before the scattering and the final state transferred to the particle by the radiation itself.
We have considered the comparative scenario in which the interaction has a finite duration, i.e., the spacetime region containing the radiation field is sandwiched between two Minkowskian zones, so that the initial state of the particle is assumed to be the same in all cases.
The final one depends instead on the properties of the different radiation fields.
In the case in which the radiation field is represented by either a photon test field in a flat spacetime or the self-consistent field of the exact electromagnetic wave, the interaction has been modeled by including a force term {\it \`a la} Poynting-Robertson into the equations of motion given by the 4-momentum density of radiation observed in the particle's rest frame with a multiplicative constant factor expressing the strength of the interaction itself.
The resulting motion is therefore not geodesic in both cases.
On the contrary, in the case in which the radiation field is represented by the gravitational field of a single plane gravitational wave, particles propagate along geodesics.

We have computed the boost (related to the simpler specific 4-momentum difference $\Delta U$) relating the initial and final $4$-momentum of the particle, $U_{(0)}$ and $U$,  both understood in the context of  the flat spacetime zones which sandwich the interaction region in between.
For the various cases and with the notation considered above, we have found for the projection of $\Delta U$ on the transverse $x$-$y$ plane
\begin{eqnarray}
\Delta U_{\rm (tf)}^\perp &=& \left(e^{-\zeta_{\rm (tf)}(u_1)}-1\right)\left(p_x\partial_x+p_y\partial_y\right)\,,\nonumber\\
\Delta U_{\rm (gw)}^\perp&=&p_x[\cos(b_{\rm (gw)} u_1)-1]\partial_x+p_y[\cosh(b_{\rm (gw)} u_1)-1]\partial_y\,,\nonumber\\
\Delta U_{\rm (em)}^\perp&=&\left(e^{-\zeta_{\rm (em)}(u_1)}\cos(b_{\rm (em)} u_1)-1\right)(p_x\partial_x+p_y\partial_y)\,,
\end{eqnarray}
whereas for the projection  on the transverse $u$-$v$ plane 
\begin{eqnarray}
\Delta U_{\rm (tf)}^\Vert &=&-\frac{1}{2p_v}\left[\left(e^{\zeta_{\rm (tf)}(u_1)}-1\right)+p_\perp^2\left(e^{-\zeta_{\rm (tf)}(u_1)}-1\right)\right]\partial_v
-p_v\left(e^{-\zeta_{\rm (tf)}(u_1)}-1\right) \partial_u\nonumber\\
\Delta U_{\rm (gw)}^\Vert&=&-\frac{1}{2p_v}[-p_x^2 \sin^2(b_{\rm (gw)} u_1)+p_y^2\sinh^2(b_{\rm (gw)} u_1)]\partial_v \nonumber\\
\Delta U_{\rm (em)}^\Vert&=&
-\frac{1}{2p_v}\left[\left(e^{\zeta_{\rm (em)}(u_1)}-1\right)+p_\perp^2\left(\cos^2(b_{\rm (em)}u)e^{-\zeta_{\rm (em)}(u_1)}-1\right)\right]\partial_v
-p_v\left(e^{-\zeta_{\rm (em)}(u_1)}-1\right)\partial_u\,.
\end{eqnarray}
In the limit of small electromagnetic field compared to the duration of the wave $|b_{\rm(em)}|u_1=\sqrt{2}|\omega_{(\rm em)}|u_1  \ll 1$ 
in the final case of  the exact electrovac solution field, one obtains the same result as in the first case of a test field with the same frequency $\omega_K=\omega_{\rm(em)}$ and therefore the same radiation field energy-momentum tensor.
In the gravitational case the transverse change in the momentum involves a rotation due to the deformation of the plane wave directions by the wave, while in the electromagnetic cases only an overall scaling is involved. For the longitudinal changes, the gravitational case lacks a component along $\partial_u$ because the motion is geodesic and $\partial_v$ is a Killing vector field, while in the other cases the force responsible for the change in momentum itself has a covariant component along $\partial_v$.

This comparative analysis shows how the nature of the interaction of massive particles with radiation fields of different kind strongly influences the scattering process, in principle leading to detectable observational consequences.

\begin{acknowledgments}
The authors acknowledge ICRANet for support. 
\end{acknowledgments}


\begin{thebibliography}{00}


\bibitem{grif3}
J. B. Griffiths, 
{\it Colliding Plane Waves in General Relativity} (Oxford University Press, Oxford, 1991).

\bibitem{tsagas}
C. G. Tsagas,
Phys. Rev. D {\bf 84}, 043524 (2011).

\bibitem{garriga}
J. Garriga and E. Verdaguer,
Phys. Rev. D {\bf 43}, 391 (1991).

\bibitem{bfho}
D. Bini, P. Fortini, M. Haney and A. Ortolan,
Class. Quantum Grav. {\bf 28}, 235007 (2011).

\bibitem{bgscat}
D. Bini and A. Geralico,
Phys. Rev. D {\bf 85}, 044001 (2012).

\bibitem{Poynting-03}
J. H. Poynting, 
Phil. Trans. R. Soc. A {\bf 202}, 525 (1904).

\bibitem{Robertson-37}
H. P. Robertson,  
Mon.\ Not.\ R.\ Astron.\ Soc. {\bf 97}, 423 (1937).

\bibitem{BiniJS-09}
D. Bini, R. T. Jantzen, and L. Stella,
Class.\ Quantum Grav. {\bf 26}, 055009 (2009).

\bibitem{BiniGJSS-11}
D. Bini, A. Geralico, R. T. Jantzen, O. Semer\'ak, and L. Stella, 
Class.\ Quantum Grav. {\bf 28}, 035008 (2011).

\bibitem{vaidyaPR}
D. Bini, A. Geralico, R. T. Jantzen, and O. Semer\'ak,
Class.\ Quantum Grav. {\bf 28}, 245019 (2011).

\bibitem{Vaidya-43}
P. C. Vaidya,
Current Sci. (India) {\bf 12}, 183 (1943).

\bibitem{grif2}
J. B. Griffiths, 
Ann. Phys. (N.Y.) {\bf 102}, 388 (1976).

\bibitem{rindler}
Rindler W 2001
 \textit{Relativity:
special, general, and cosmological}
(Oxford University Press) 

\bibitem{bp}
Bondi H, Pirani F A E and Robinson I 1959
{\it Proc.\ Roy.\ Soc. London\/} {\bf 251} 519

\bibitem{grif}
J. B. Griffiths, 
Phys. Lett. A {\bf 54}, 269 (1975).



\end{thebibliography}
\end{document}